# Readout Electronics of T0 Detector in the External Target Experiment of CSR in HIRFL

Peipei Deng, Lei Zhao, *Member*, IEEE, Jiaming Lu, Pinzheng Xia, Jinxin Liu, Min Li, Shubin Liu, Member, IEEE, and Qi An, Member, IEEE

*Abstract*–T0 detector, based on Multi-gap Resistive Plate Chambers (MRPC) technology, is one of the key components in the External Target Experiment. Through precision measurements of the MRPC signals, timing of the beam impact on target can be obtained and used as the start time for other detectors. A readout electronics system was designed for the T0 detector. Based on the NINO ASIC, front-end-electronics (FEE) circuits which can achieve high precision leading-edge discrimination and Charge-to-Time Conversion (QTC) were designed for the internal and external MRPCs of the T0 detector. The output pulse of the FEE is then digitized by high precision time digitization modules with Time-to-Digital Converters (TDCs), trigger matching and other control logic integrated within Field Programmable Gate Array (FPGA) devices. To evaluate the functionality and performance, we also conducted a series of tests of the electronics. Results indicate that the system functions well and the time precision of the electronics is better than 21 ps, which satisfies the application requirement.

## I. INTRODUCTION

HEAVY Ion Research Facility in LanZhou (HIRFL) is the largest heavy ion research facility in China. The Cooling Storage Ring (CSR) project in HIRFL consists of a main ring (CSRm), an experiment ring (CSRe), and a radioactive beam line (RIBLL2) [1, 2]. The External Target Experiment in CSR is dedicated for heavy ion collision research [3]. It consists of multiple types of detectors, as shown in Fig. 1, including a start time (T0) detector, a γ detector, a big dipole, six Multi-Wire Drift Chambers (MWDCs), three Time of Flight (TOF) walls, and a neutron wall. The T0 detector is located in the front of all other detectors in the system and provides accurate start time information for the entire system. After an ion beam bunch impinges on the target, other detectors record the properties of the particles scattered after collision. The repetition rate of the bunch is about half a minute, and the duration time of each bunch is 4 seconds. Since there are less than $10^6$ charged particles in each bunch and around 1% of the particles interact with the target, there are less than 2500 particle tracks generated per second.

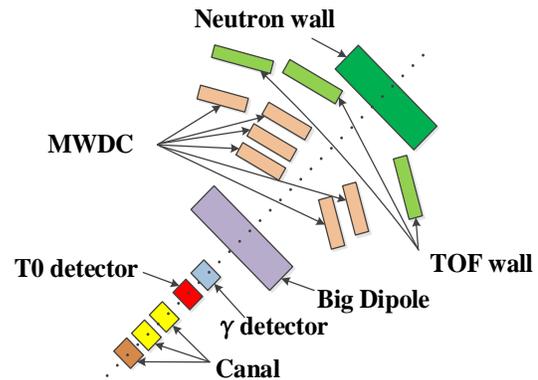

Fig. 1. Detectors of the CSR external target experiment in HIRFL.

The architecture of the T0 detector is shown in Fig. 2. The detector consists of internal and external Multi-Gap Resistive Plate Chambers (MRPCs). In order to reduce deterioration of the MRPC signal, the Front-End-Electronics (FEE) modules are placed in close proximity of the MRPC detectors. Digitization modules are connected to the FEE through ten-meter cables. In the first phase, two internal and two external detectors are implemented, corresponding to 80 electronics channels. In the next phase, six internal and six external detectors will be implemented with 250 readout channels in total. For each channel, the input average hit rate is 1 kHz, and the minimum time interval between two adjacent pulses is around 10 μs.

This work was supported in part by the Knowledge Innovation Program of the Chinese Academy of Sciences under Grant KJCX2-YW-N27, in part by the National Natural Science Foundation of China under Grant U1232206, and in part by the CAS Center for Excellence in Particle Physics (CCEPP).

The authors are with the State Key Laboratory of Particle Detection and Electronics, University of Science and Technology of China, Hefei 230026, China, and also with Department of Modern Physics, University of Science and Technology of China, Hefei 230026, China (Corresponding author: Lei Zhao, e-mail: zlei@ustc.edu.cn).

© 2018 IEEE. Accepted version for publication by IEEE. Digital Object Identifier 10.1109/TNS.2018.2834426

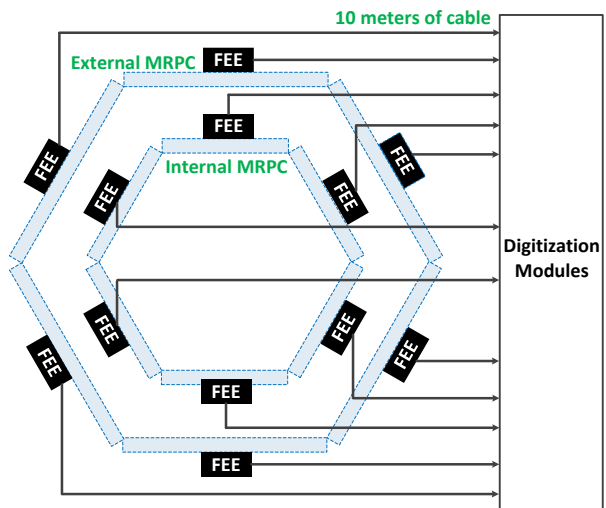

Fig. 2. Architecture of the T0 detector.

The MRPC output signal consists of differential pairs with fast leading and trailing edges. Fig. 3 shows the MRPC waveforms after amplification.

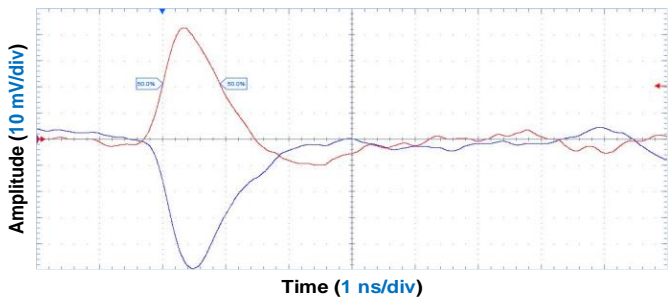

Fig. 3. Signal waveform of MRPC after amplification by Radio Frequency amplifier GALI6+ from Mini Circuits Corporation.

Two main designs of the readout electronics for MRPC detectors exist. In certain large scale experiments, such as in ALICE [4] and BESⅢ ETOF [5], the electronics are based on the NINO ASIC for amplification and discrimination, and the High Performance general purpose Time to Digital Converter (HPTDC) ASIC for time digitization. In small scale experiments or test systems for MRPC performance evaluation [6], the electronics are based on commercial discrete devices for amplification and the HPTDC board for digitization. In the T0 detector, readout electronics are based on the NINO ASIC for the FEE and the FPGA-based Time-to-Digit Converter (TDC) for digitization.

## II. SYSTEM ARCHITECTURE

The readout electronics system of T0 detector (in Phase I) is shown in Fig. 4, consisting of four FEE modules, four time-digitization Modules (TDMs), one power module, one Slave Trigger Module (STM), one Slave Clock Module (SCM), one Master Trigger Module (MTM) and one Master Clock Module (MCM). Several of the above modules have previously been designed for other detectors in the External Target Experiment, e.g. the clock system including SCM and MCM [7], which provides high quality clock signals distributed to different measurement modules, as well as the trigger system including STM and MTM [8], which are essential for valid data readout and background noise rejection. This paper focuses on the design of the FEE and TDM, which were specially designed for T0 detector.

The FEE, based on the TOT (Time Over Threshold) method, processes analog signals from the MRPC detector and outputs Low Voltage Differential Signals (LVDS). The LVDS signals are then digitized by the TDC modules, based on the PXI (PCI extensions for Instrumentation) [9] 6U standard, in order to obtain the leading edge and pulse width. Subsequently, the measurement data are transferred to the PC through the PXI interface. Readout electronics in Phase I consists of two FEEs for internal MRPC detectors, two FEEs for external MRPC detectors and four TDMs.

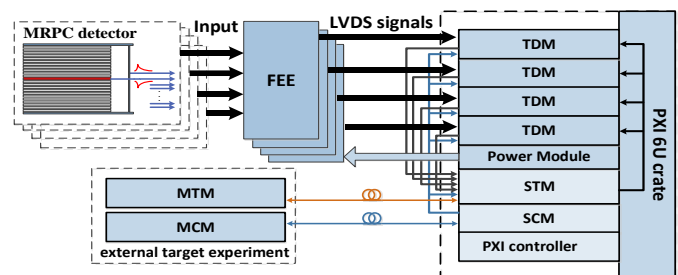

Fig. 4. Architecture of the readout system.

## III. FRONT-END-ELECTRONICS DESIGN

In order to meet the stability and performance requirements, we employ the NINO ASIC [10, 11] for signal amplification. Based on the NINO ASIC, two different FEE modules were designed for internal and external MRPC of the T0 detector. Each internal FEE module includes two NINO ASICs, and implements 16-channel analog signal processing. Similar to the internal FEE module, one external FEE module contains four NINO ASICs with 24 channels in total. The architecture of the FEE is shown in Fig. 5. Each input channel uses a Transient Voltage Suppressors (TVS) diode for overcharge protection. Analog signals from MRPC detectors are processed by the NINO chips, which are fully differential integrated circuits. At the input stage, an off-chip resistor is utilized for impedance matching. After that, the signal is amplified by four identical cascaded amplifiers. Furthermore, there is a pulse stretcher in front of the LVDS output driver. The pulse width before stretching varies between 2 ns and 7 ns and is increased by 10 ns after stretching. In addition, a logical OR operation can be performed on the 8 channel output signals before output stage, and the resulting signal can be used as a simple trigger signal. The photographs of internal and external FEE module are shown in Fig. 6.

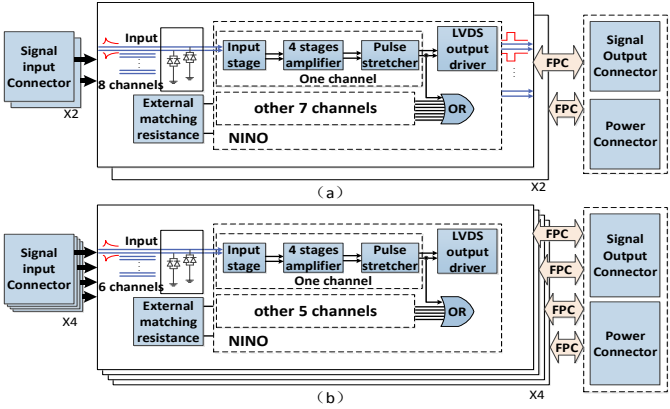

Fig. 5. Architecture of the FEE.

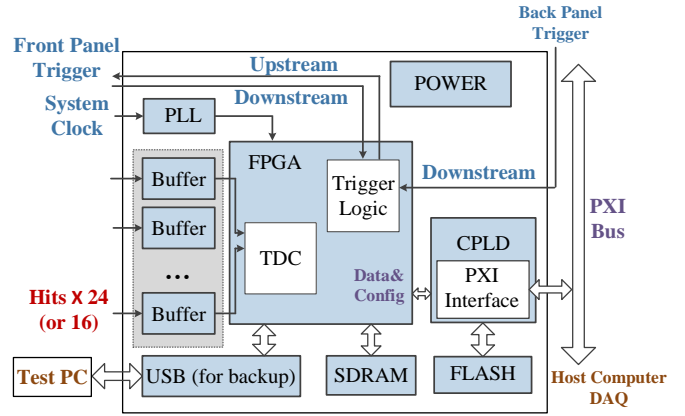

Fig. 8. Diagram of the Time Digitization Module.

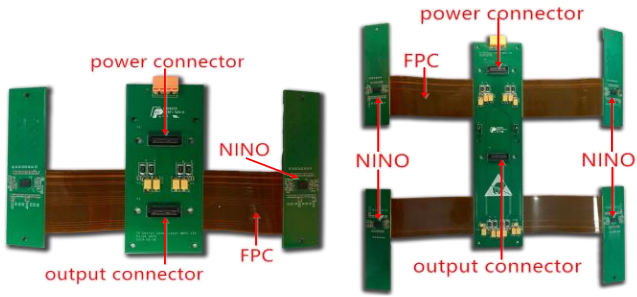

Fig. 6. Internal (left) and external (right) FEE modules.

## IV. TIME DIGITIZATION MODULE DESIGN

The TDM (shown in Fig. 7), implemented on a Xilinx Artix-7 series FPGA, has 24 TDC channels, trigger preprocessing logic, trigger matching logic and host computer communication function.

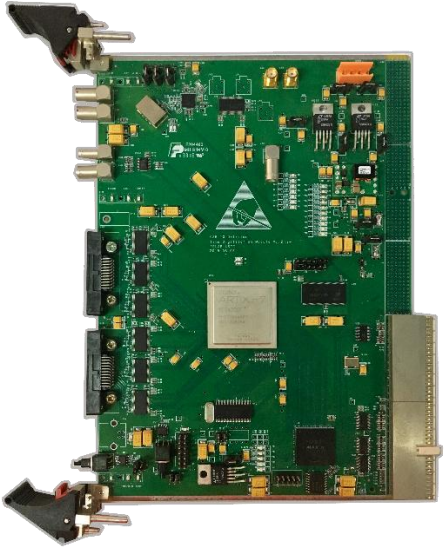

Fig. 7. Photograph of Time Digitization Module.

The block diagram of TDM is presented in Fig. 8. 24-channel TOT signals for external MRPC detectors and 16-channel for internal MRPC detectors are transmitted from the FEE into the TDM through high-density coaxial cables. Subsequently, they are buffered and transferred into the FPGA for time measurement. After TDM receives the clock signal from the clock system, it becomes synchronized with other measurement modules. The Phase Locked Loop (PLL) is used for jitter cleaning and frequency conversion, and it also outputs a high-quality clock signal which drives FPGA internal logic. Trigger signal fan-out from the SCM can be connected either through a single-ended LEMO connector, or the start trigger bus in the backplane of the PXI crate. The PXI bus interface is implemented inside the Complex Programmable Logic Device (CPLD), which receives the data from the FPGA and then transfers measurement results to the Single Board Computer (SBC) in the Slot 0 of the crate, and finally to the DAQ. In addition, the FPGA logic is stored in a FLASH memory connected to the CPLD and can be updated online by a remote PC to meet the practical requirements of the experiment.

### A. TDC Logic Design

The advantages of using FPGA to implement TDC functionality are good flexibility and system simplicity. Since the TDC is implemented in the FPGA, the parameters can be customized and changed as needed. In addition to TDC logic, other controlling logic circuits can be integrated within a single FPGA device. At present, most commonly implemented methods are multiphase clock interpolation [12], pulse shrinking [13], Tapped Delay Line (TDL) [14], the Vernier method [15], and the Vernier delay line (VDL) method [16]. In our case, the TDL method is used in order to achieve the required time resolution of 25 ps RMS.

The structure of TDL used here is based on the carry chain inside the FPGA slice resource, as shown in Fig. 9. In the carry chain of each slice, there are four buffers (called CARRY-FOUR), whose output is connected to a D flip-flop. The CARRY-FOURs are connected in series, allowing construction of a delay chain with a large dynamic range.

TOT signals. This can be achieved in a single carry chain to reduce the cost instead of implementing two chains and measuring them separately [17]. In Fig. 9, CARRY4 contains 4 delay taps with 4 outputs, CO1 - CO4 that are connected to the inputs of the D flip-flops via multiplexers. There are two types of signal paths connected to each output of the tap. One (CO1) is connected to the multiplexer and finally to the D flip-flop, while the other (CO3) signal path is connected to the D flip-flop through a XOR gate and a multiplexer. The second input of the XOR gate is connected to the high level '1', and thus the trailing edges are converted to leading edges and then recorded by the D flip-flops. Fig. 10 shows the encoding scheme for the two sets of thermometer codes, controlled by the leading and trailing edge detection logic.

## B. Trigger Logic Design

As shown in Fig. 11, the trigger preprocessing logic collects the signals of each TDC channel and determines whether it agrees with the expected mode. Trigger matching logic receives the trigger signal from SCM, and extracts the valid data in the pre-set matching window. During the experiment, the trigger rate of the T0 detector is less than 10 kHz. The trigger latency can be configured (200 ns is used in the experiment), with a maximum value of 2000 ns. The trigger window width is also configurable and set to 100 ns in this case.

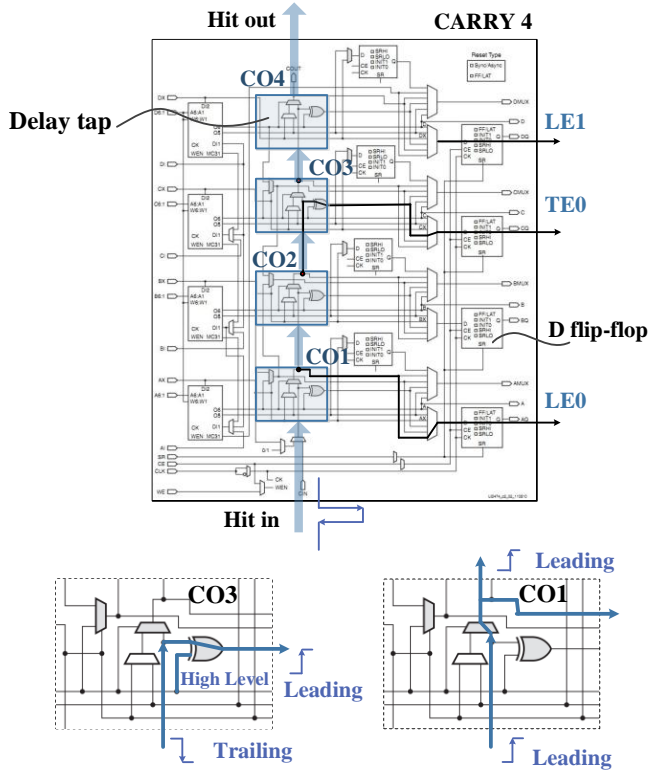

Fig. 9. Diagram of CARRY-FOUR and the realization of the measurements of leading and trailing edges in a single carry chain.

As shown in Fig. 10, 70 CARRY-FOURs with a total delay time longer than one coarse count clock (240 MHz) period (according to the simulation result) are used to construct the delay chain. When the input hit signal is transmitted along the delay chain, fine time information can be obtained in the form of thermometer code by the D flip-flop array that records the level state on the delay chain at a certain time. The latched code is subsequently encoded to the fine time measurement result. The final result, consisting of the encoded 8-bit fine time and the coarse time results from the 240 MHz-clock-driven 36-bit counter is stored into a channel FIFO.

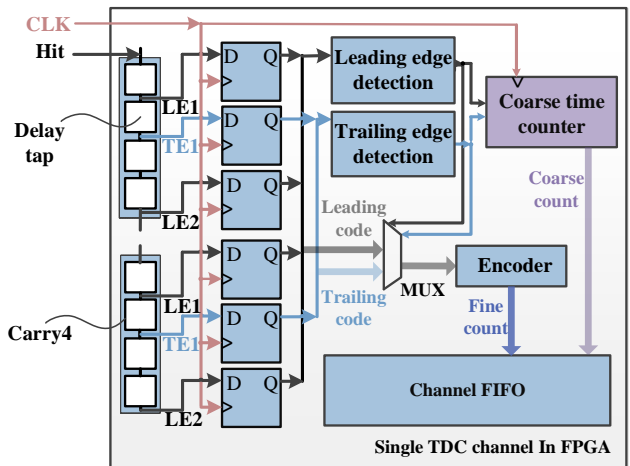

Fig. 10. Diagram of the encoding scheme for the two sets of thermometer codes.

For time-walk error correction, the TDC is required to be capable of measuring both leading and trailing edges of the

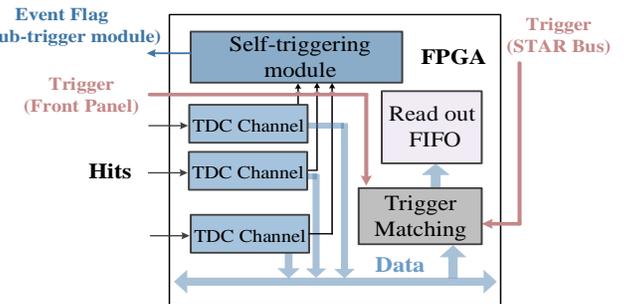

Fig. 11. Trigger functionality in the FPGA.

### 1) Trigger Preprocessing Logic

Once the hits in an event agree with the preset trigger pattern, a valid event flag signal is sent to the SCM via a dedicated trigger cable. The trigger modes of internal and external MRPC detectors are different. Regarding the internal detector (Fig. 12), whenever a particle flying from the target hits the MRPC detector, the event is considered to be valid, and an output signal will be generated and sent to SCM.

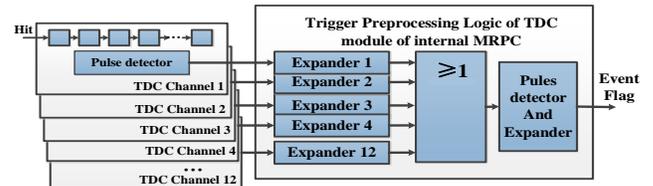

Fig. 12. Diagram of Trigger Preprocessing Logic for internal MRPC.

Regarding the external detector (Fig. 13), an output signal is generated and sent out when the signals are detected simultaneously at both ends of any of the MRPC strips.

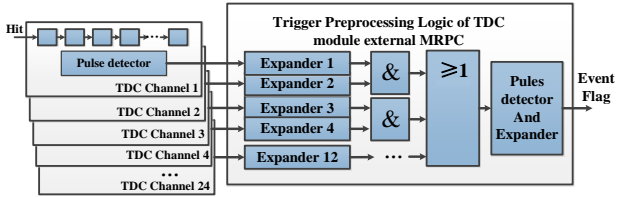

Fig. 13. Diagram of Trigger Preprocessing Logic for external MRPC.

*2) Trigger matching Logic*

Fig. 14 shows the trigger matching function in the TDM. The Trigger Info block records a variety of trigger data useful for window extraction and data packing. The address accumulator block provides round-robin write addresses for cyclic writes to Double Port Random Access Memory (DPRAM) and Content-Addressed Memory (CAM). Trigger matching control block controls the internal sub-module workflow. Furthermore, driven by the internal state machine, it extracts time information of the trigger signal from the Trigger Info block and calculates the low and high boundaries of the matching window. Content addressing to CAM is initiated for every time point within the matching window. Once the matching time point is found, the CAM encoder transcodes the output address to binary code and the data stored at this address are read out from the DPRAM. Together with bunch ID and Event ID (the order information of trigger signal) from the Trigger Info block, the time data of valid events are then sent into the readout FIFO.

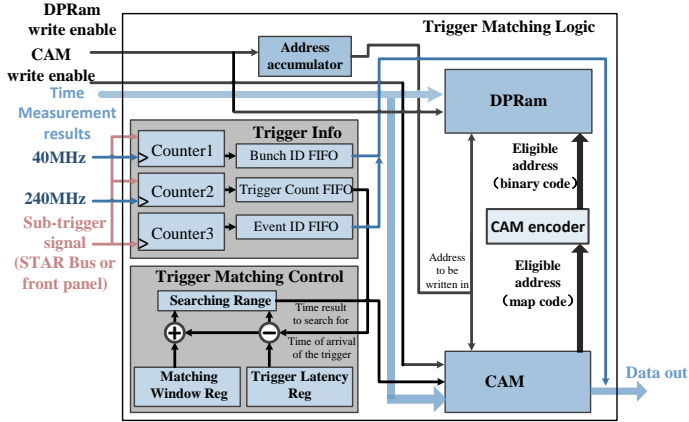

Fig. 14. Diagram of the Trigger Matching Logic.

Using CAM, a data address search can be accomplished in a single clock cycle, facilitating the trigger-matching process. Xilinx Corporation provides open source logic code, making it straightforward to customize CAM blocks with different depth and bit widths. One example of using CAM to find valid data is shown in Fig. 15. During the writing process, data and the address are received from the DIN and WR_ADDR ports, respectively. If the WRITE EN signal is enabled, input data are written to the corresponding addresses. During the process of readout, if there is no corresponding result stored in CAM, the SINGLE_MATCH and the MULTI_MATCH ports output '0'. If there is at least one corresponding result in CAM, the MATCH port outputs '1'. SINGLE_MATCH and MULTI_MATCH ports indicate the number of addresses that have the same time measurement data stored. CAM only stores coarse time information of the events while RAM stores both the coarse and fine time results, which means that a single time result stored in CAM may correspond to two or more events in different addresses in RAM. In the example shown in Fig. 15 (a), the MATCH_ADDR outputs the address '01000000' (the MSB is the leftmost bit), which represents the '111' binary coded address in RAM. For this condition, SINGLE_MATCH outputs '1' and MULTI_MATCH outputs '0'. In Fig. (b), MATCH_ADDR outputs the address '00100001', representing the two addresses that correspond to '110' and '001' address of the RAM. SINGLE_MATCH outputs '0' and MULTI_MATCH outputs '1'.

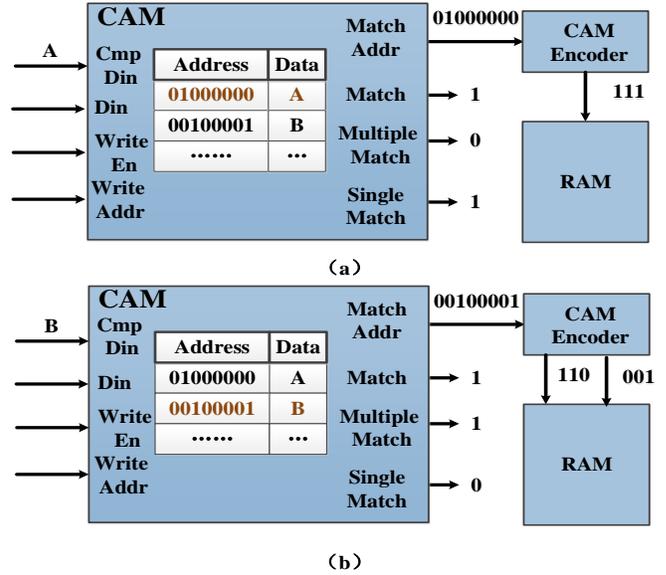

Fig. 15. Example of the application of CAM for a valid data search.

### C. Clock Circuits

Since high precision time measurement is required, a high-quality clock circuit is needed in the TDMs. The clock circuits in each TDM are shown in Fig. 16. The 40 MHz system clock signal received from the SCM is split and one of the resulting signals connected to a Dual Loop PLL LMK04031 [18] from the Texas Instruments Corporation, with an external Voltage Controlled X-tal Oscillator (VCXO) CVPD-920 from Crystek Corporation [19], in order to produce a 240 MHz clock, which is fed to the FPGA. The other clock signal directly enters the FPGA, as a backup. The Digital Clock Manager (DCM) block inside the FPGA then generates multiple clocks for the entire FPGA logic.

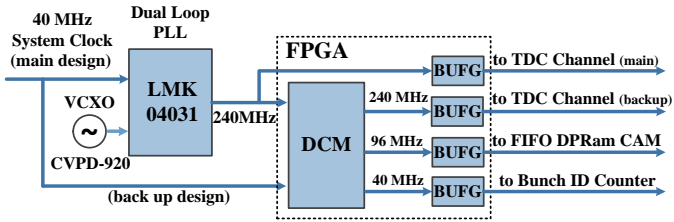

Fig. 16. Architecture of the clock for digitization module.

## D. Data Buffering and Transfer

Multi-level token ring structure is used in the TDM in order to transfer multi-channel measurement data to a common readout FIFO (shown in Fig. 17), as well as for PXI data readout.

Since there are multiple TDC channels in the TDM, in certain extreme situations a large amount of data could be stored within one channel, with only minute amounts of data in the other channels. However, the token ring structure guarantees that each channel can access the common data path with equal priority.

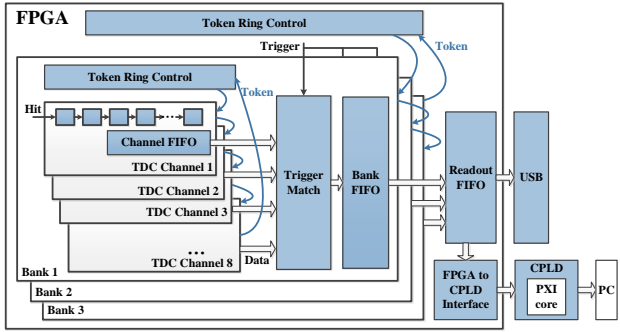

Fig. 17. Diagram of data transfer processing.

24 TDC channels are divided into 3 banks (for external MRPCs; there are only 2 banks for the 16 TDC channels for internal MPRCs). Each bank contains 8 channels, a trigger matching block, and a bank FIFO. In the first step of data transfer, measurement data from the Channel FIFO is sent to the trigger matching block, where the first-level token ring is executed. Each bank contains token ring control logic (Token Ring Control), and each channel contains token processing logic. Token ring control block receives the "Empty" signals from the eight Channel FIFOs and initiates token ring transmission. The token processing logic in each channel receives the output token from the previous channel, transfers all data to the trigger matching block, and then releases the token to the next channel. The last channel returns the token to the token ring control block and becomes ready for the next iteration. Data transfer from the banks to the readout FIFO is based on similar token ring structure.

## V. TEST RESULTS

In order to evaluate the function and performance of the electronics system, we conducted a series of laboratory tests. Fig. 18 shows the system under test.

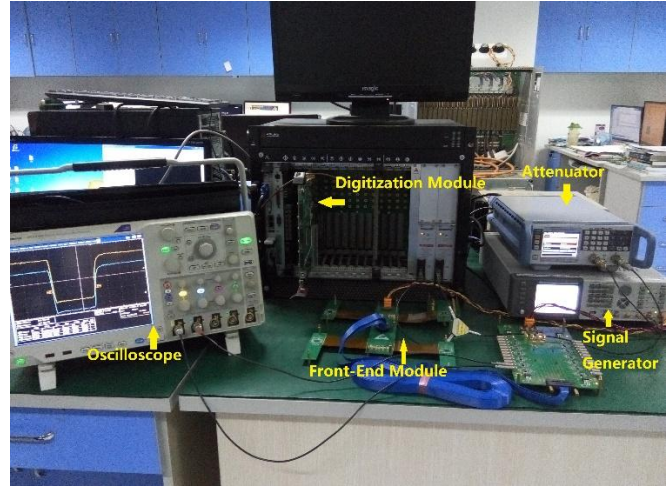

Fig. 18. Photograph of the test system.

## A. FEE Performance

Internal FEE test results are shown in Fig. 19. The time resolution of all 32 channels for the internal FEEs and the 48 channels for the external FEEs are all better than 10 ps RMS in the dynamic range from 400 fC to 2 pC.

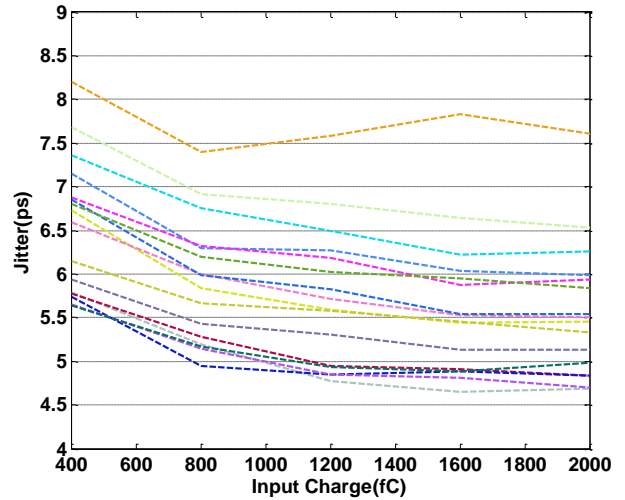

(a)

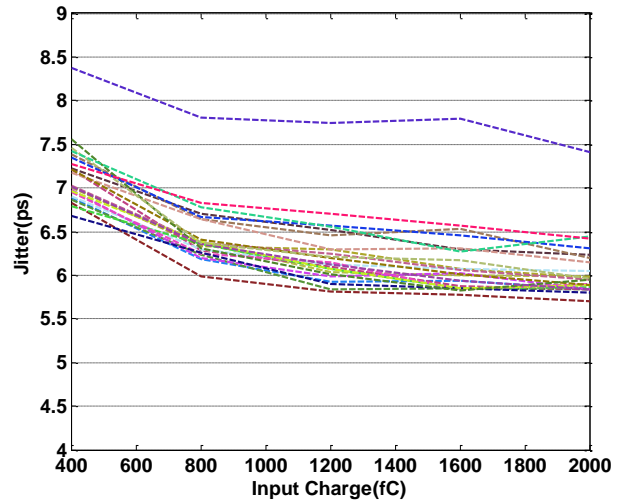

Fig. 19. Time resolution test results. The (a) shows internal MRPC, and (b) external MRPC results. The input charge is varied from 400fC to 2 pC.

Fig. 20 shows the pulse widths of internal and external FEEs, which agree well with expectations.

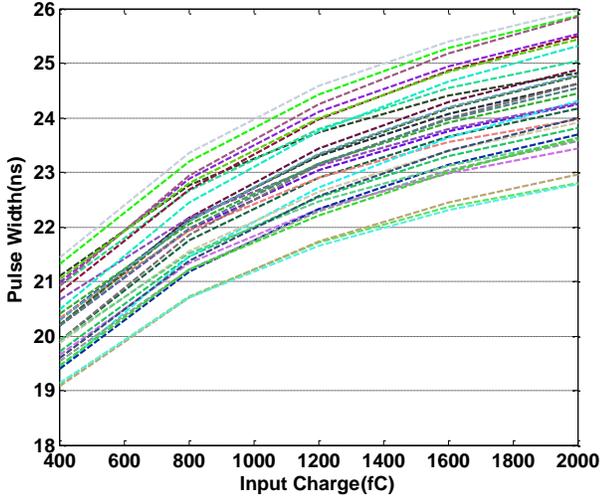

(a)

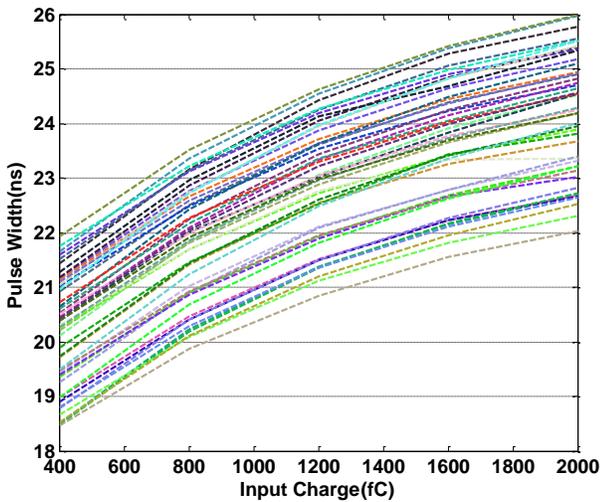

(b)

Fig. 20. Pulse widths test results. The (a) is for internal MRPC, and (b) is for external MRPC. The input charge is swept from 400fC to 2 pC.

## B. TDM Performance

Fig. 21 shows the Integral Non-Linearity (INL) test results for the leading edge of a typical channel. The INL varies from -1.3 to 2.2 Least Significant Bit (LSB).

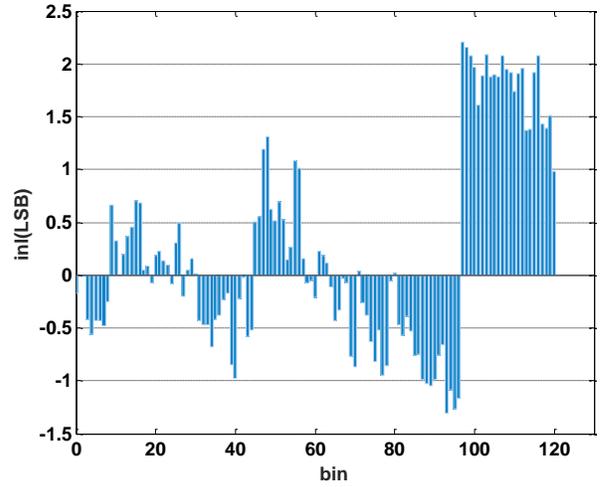

Fig. 21. Non-Linearity test of the leading edge of one typical channel.

Fig. 22 and 23 show the time resolution results of the two TDMs designed for the internal MPRC, while Fig. 24 and 25 show the two TDMs for the external MRPC. As can be observed in these figures, all of the 80 channels exhibit time resolution ranging from 15 ps to 22 ps RMS.

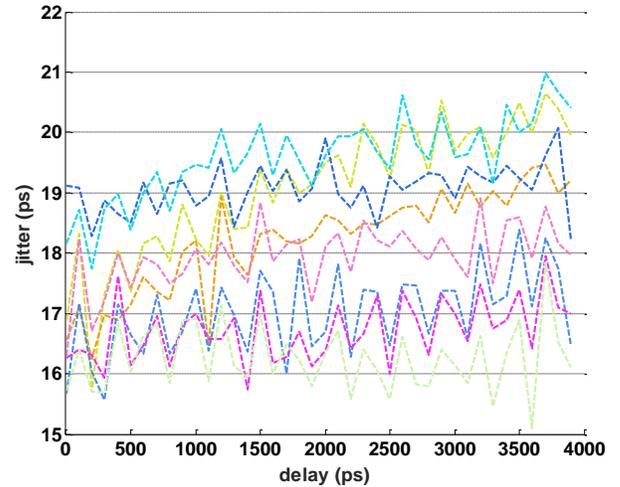

Fig. 22. TDM 1 time resolution.

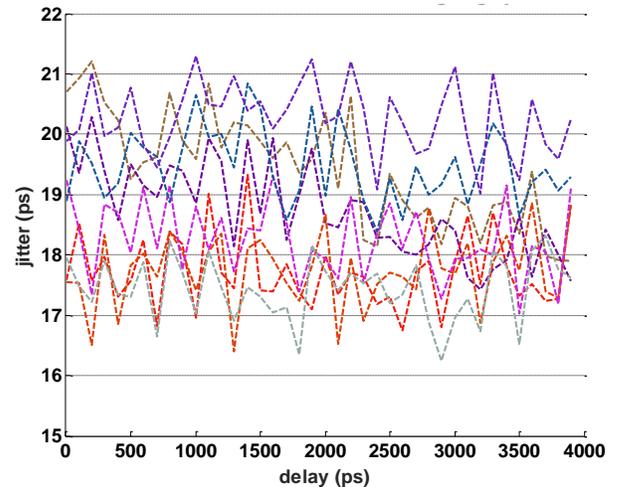

Fig. 23. TDM 2 time resolution.

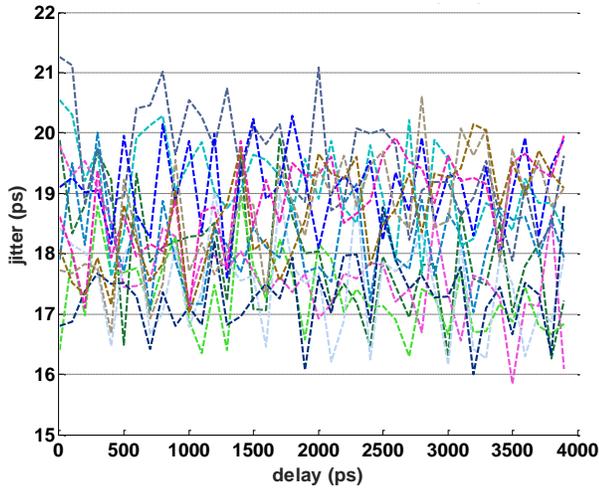

Fig. 24. TDM 3 time resolution.

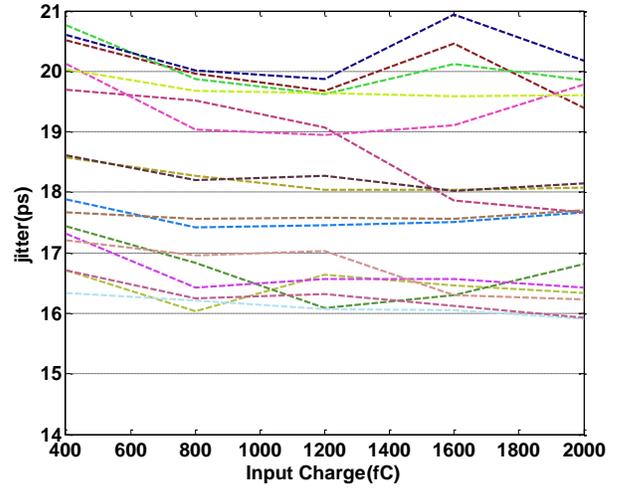

(a)

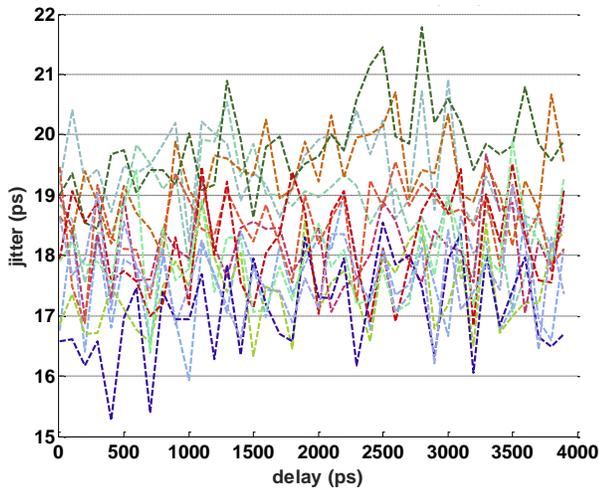

Fig. 25. TDM 4 time resolution.

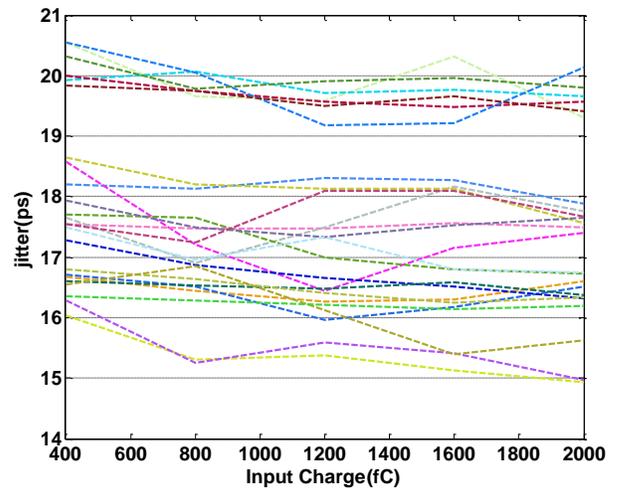

(b)

Fig. 26. Time resolution test results of the electronics.

## C. Performance of the entire electronic system

Fig. 26 shows the overall time resolution of the electronics (including both the FEEs and TDMs), and Fig. 26 (a) and (b) show the results for internal and external MRPCs, respectively. The time resolution of each of the 80 channels is better than 21 ps in the dynamic range from 400 fC to 2 pC, which meets the requirements for this application.

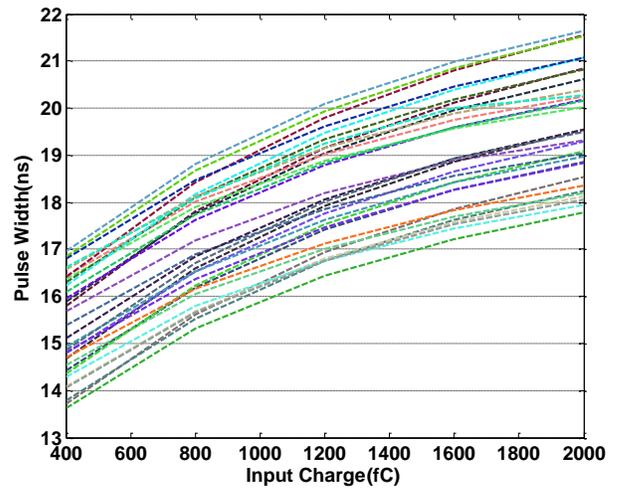

(a)

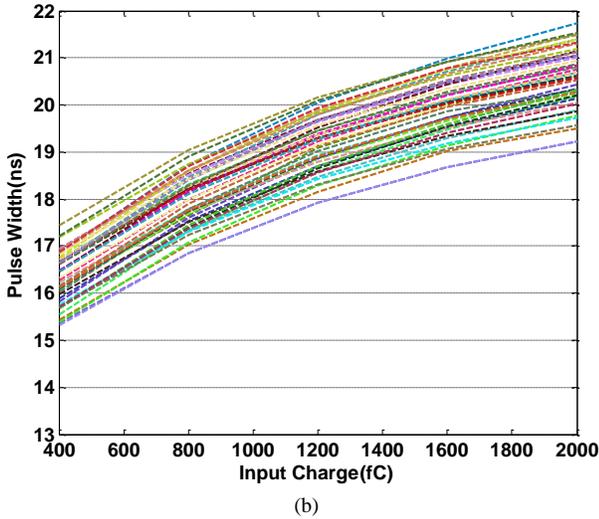

Fig. 27. Pulse widths test results of the electronics. The input charge is swept from 400 fC to 2 pC.

Fig. 27 shows the pulse widths test results for the overall electronics, and Fig. 27 (a) and (b) correspond to the internal and external MRPCs, respectively. The results of all 80 channels are in agreement with expectation.

### D. Test results of the MRPC Detector combined with electronics

The diagram of the test system is shown in Fig. 28. The signals from the MRPC are fed into FEE and recorded by the TDM, while the signals from the four PMTs are measured by HPTDCs. The trigger given to the system is generated by the coincidence of the signals of the four PMTs, and the average timing of the four PMTs are used as the reference time (T0, Fig. 29, fitted with a Gaussians function with a variance (σ) of 72.1 ps.).

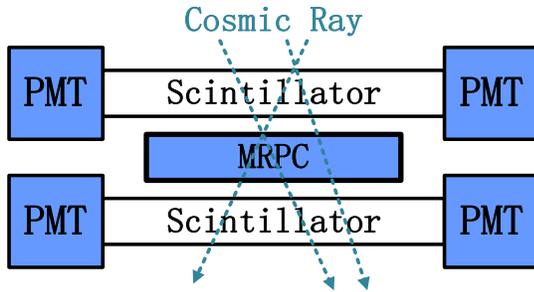

Fig. 28. Diagram of the MRPC detector combined with the electronics test system.

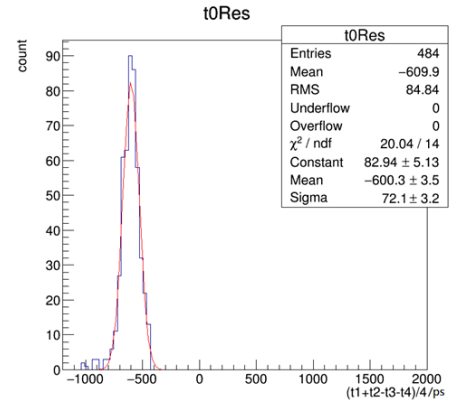

Fig. 29. T0 distribution.

A typical MRPC pad time distribution relative to T0, corrected for the T-TOT correlation effect, is shown in Fig. 30. It has a variance (σ) of 108.8 ps.

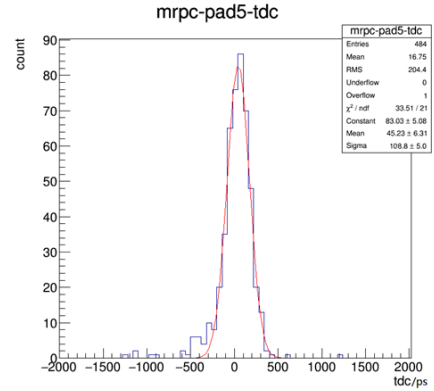

Fig. 30. A typical MRPC pad time distribution

The resolution of the MRPC detector combined with electronics is characterized by the following equation:

$$\sigma = \sqrt{\sigma_t^2 - \sigma_{t_0}^2} = \sqrt{108.8^2 - 72.1^2} = 81.48 \text{ps} \quad (1),$$

which also meets our requirements.

## VI. CONCLUSION

High precision time measurement electronics were designed for the MRPC in T0 Detector of the External Target Experiment for CSR in HIRFL. Employing an FEE design based on the NINO ASIC, time and charge information can be digitized simultaneously, simplifying the system structure. An FPGA-based TDC is used in this work, integrating trigger matching and other control logic within one FPGA device in order to enhance its flexibility. Test results of the electronics indicate that the overall time resolution of the electronics is better than 21 ps RMS, which satisfies the requirements for this application.

ACKNOWLEDGMENT

We thank Xiaoshan Jiang and Hongliang Dai in Institute of High Energy, CAS for their constant help in our FEE design

work. And we also thank Ming Shao, Yongjie Sun, Dongdong Hu and Jian Zhou for their kind help in our research.